# Analysis of Birth weight using Singular Value Decomposition

D.Nagarajan
Department of Mathematics,
Salalah College of Technology,
Salalah,Sultanate of Oman.

V.Nagarajan
Department of Mathematics,
S.T.Hindu College,Nagercoil,
Kanyakumari,Tamil Nadu,India.

P.Sunitha
Department of Mathematics,
S.T.Hindu College,Nagercoil,
Kanyakumari,Tamil Nadu,India

V.Seethalekshmi
Department of Mathematics,
James College Technology,
Nagercoil,TamilNadu,India.

*Abstract*— **The researchers have drawn much attention about the birth weight of newborn babies in the last three decades. The birth weight is one of the vital roles in the baby's health. So many researchers such as [2],[1] and [4] analyzed the birth weight of babies. The aim of this paper is to analyze the birth weight and some other birth weight related variable, using singular value decomposition and multiple linear regression.**

## I. INTRODUCTION

For a successful reproduction a good health throughout childhood, adolescence adult life and pregnancy is necessary. Special care has to be taken during pregnancy to get a healthy baby. Birth weight is an important determination of child health. It is a well known fact that the birth weight in influenced by the factors such as gestational period, maternal height, maternal weight, age, parity, haemoglobin concentration, rate of intrauterine growth, nutrition and many other socio-economic factors. Low birth weight neonate is defined as any neonate weighing less than 2500 grams at birth. It raises grave health risks for children which is a public health problem in most developing countries. Low birth weight stems primarily from poor maternal health and nutrition. Three factors have most impact poor maternal nutritional status before conception, short stature due mostly to under nutrition and infections during child hood and poor nutrition during pregnancy .Less than 17 years and greater than 34 years of age are at increased risk of low birth weight delivery. It can arise as a result of a baby being born too earlier (<37 weeks, also known as premature birth) or being born too small for gestational age (small as a result of intrauterine growth restriction). These types of babies have worst prognosis. Since low birth weight children are responsible for a very significant proportion of morbidity and mortality in childhood whereas child born with adequate birth weight are reported to do well even under adverse environment, researchers often use birth weight as a measure of morbidity risk.

## II. DATA BASE

The data for the present study were collected from Chennai some private hospital during January 2008 to December 2008. The data for the selected variables were collected from the hospital case records.

## III. DESCRIPTION OF MODEL

The singular value decomposition closely associated to the companion theory of diagonaling a symmetric matrix. Hark back that if A is a symmetric real n x n matrix there is an orthogonal matrix V and a diagonal D such that

$$A=VDV^T. \quad (1)$$

Here the columns of V are latent vectors for A and diagonal entries of D are eigen values of A for Singular Value Decomposition begin with m x n real matrix. There are orthogonal matrices U and V and a diagonal matrix S, such that

$$A = USV^T. \quad (2)$$

Here U is m x m and V is n x n, so that S is rectangular with the same dimensions as A .The matrix S can be formatted to be non negative and in order of decreasing order. The columns of U and V are called left and right singular vectors for A.[3]. Singular value decomposition is used in Latent Semantic Indexing (LSI) to determine the rank of the maternal variables and birth weight. Before scoring the maternal variables with Latent Semantic Indexing we need to onstruct a matrix with the maternal variables available as "A" .

TABLE I : MATERNAL VARIABLES BY BIRTH WEIGHT

| Baby Weight | Maternal height (145-162) | Maternal Weight (50-70) | Age (21-35) | Blood Pressure (120/80) | Haemo globin (9-14) |
|---|---|---|---|---|---|
| < 2000 | 15 | 10 | 8 | 15 | 9 |
| 2000 | 78 | 55 | 58 | 35 | 60 |
| 2400 | 120 | 105 | 133 | 102 | 80 |
| 2800 | 206 | 240 | 230 | 180 | 150 |
| 3200 | 120 | 135 | 142 | 60 | 53 |
| 3600 | 8 | 21 | 15 | 25 | 19 |
| >3600 | 3 | 7 | 4 | 10 | 3 |

The "Baby weight, Y" is taken as the dependent variable and other maternal variables are treated as independent variables X1, X2, X3, X4, X5, where

X1 : Maternal Height

X2 : Maternal Weight

X3 Age of mothers

X4 : Blood pressure

X5 : Haemoglobin concentration

Y : Baby weight

According to Singular Value Decomposition theory, an arbitrarily real rectangular matrix of order m x n can be decomposed into three matrices such that

$$A_{m\times n} = U_{m\times l}\; S_{l\times n}\; V_{n\times n}^{T}\,, \qquad (3)$$

where U and V are orthogonal matrices and S is a singular matrix with eigen values as its diagonal entries ,which are arranged in non - increasing order. The following analysis using MATLAB.

$$A=\begin{bmatrix}15 & 10 & 8 & 15 & 9\\ 78 & 55 & 58 & 35 & 60\\ 120 & 105 & 133 & 102 & 80\\ 206 & 240 & 230 & 180 & 150\\ 120 & 135 & 142 & 60 & 53\\ 8 & 21 & 15 & 25 & 19\\ 3 & 7 & 4 & 10 & 3\end{bmatrix}$$

$$U=\begin{bmatrix}-0.0424 & 0.1362 & -0.0520 & 0.0642 & 0.7905 & 0.5108 & -0.2948\\ -0.2177 & 0.1121 & -0.8923 & -0.3417 & -0.0643 & 0.0783 & 0.1293\\ -0.4183 & 0.2287 & -0.1740 & 0.8457 & -0.1576 & 0.0405 & 0.0284\\ -0.7774 & 0.2715 & 0.3157 & -0.3517 & 0.1051 & -0.2749 & -0.0906\\ -0.4088 & -0.8714 & 0.0160 & 0.0182 & -0.0222 & 0.2636 & 0.0547\\ -0.0637 & 0.2748 & 0.2295 & -0.1896 & -0.5022 & 0.7609 & 0.0261\\ -0.0201 & 0.0897 & 0.1348 & 0.0114 & 0.2869 & 0.0853 & 0.9400\end{bmatrix}$$

$$S=\begin{bmatrix}585.9583 & 0 & 0 & 0 & 0\\ 0 & 49.1200 & 0 & 0 & 0\\ 0 & 0 & 34.3363 & 0 & 0\\ 0 & 0 & 0 & 21.2226 & 0\\ 0 & 0 & 0 & 0 & 6.6886\\ 0 & 0 & 0 & 0 & 0\\ 0 & 0 & 0 & 0 & 0\end{bmatrix}$$

$$V=\begin{bmatrix}-0.4737 & -0.1618 & -0.6429 & 0.1375 & 0.5632\\ -0.5133 & -0.2822 & 0.4454 & -0.6588 & 0.1564\\ -0.5230 & -0.3830 & 0.1032 & 0.5093 & -0.5565\\ -0.3705 & 0.6792 & 0.4333 & 0.3596 & 0.2903\\ -0.3182 & 0.5349 & -0.4358 & -0.3979 & -0.5142\end{bmatrix}$$

From S matrix shows that its consists of five non zero singular values, confirming that A is a rank 5 matrix.

*A. Dimensionality Reduction:*

Computing $U_K, S_K, V_K$ from U,S,V using MATLAB.

Let us take the economic dimension $K$ =2 ,that is rank 2 approximation that means the first 2 columns of U and V and the first two rows and columns of S.

*B. Decomposed matrices*

Dimensionality reduction has been done my truncating the three matrices obtained form Singular Value Decomposition

$$U_K=\begin{bmatrix}-0.0424 & 0.1362\\ -0.2177 & 0.1121\\ -0.4183 & 0.2287\\ -0.7774 & 0.2715\\ -0.4088 & -0.8714\\ -0.0637 & 0.2748\\ -0.0201 & 0.0897\end{bmatrix} \qquad S_K=\begin{bmatrix}585.95 & 0\\ 0 & 49.12\end{bmatrix}$$

$$V_K=\begin{bmatrix}-0.4737 & -0.1618\\ -0.5133 & -0.2822\\ -0.5230 & -0.3830\\ -0.3705 & -0.6792\\ -0.3182 & -0.5349\end{bmatrix} \qquad S_K^{-1}=\begin{bmatrix}0.0017 & 0\\ 0 & 0.02035\end{bmatrix}$$

It has been done by choosing the rank to be 2 ie, K = 2 is applied .Using

$$X_i = X_i^T\, U_K\, Sk^{-1}, \qquad (4)$$

and

$$Y = Y^T\, U_k Sk^{-1}\,, \qquad (5)$$

we get the new coordinate of vectors in this reduced space the new set of coordinate vectors are given below

$$Y = [-8.9121 \quad 7.4489]$$
$$X_1 = [-0.4709 \quad -0.1620]$$
$$X_2 = [-0.5107 \quad -0.2826]$$
$$X_3 = [-0.5205 \quad -0.3836]$$
$$X_4 = [-0.3681 \quad 0.6808]$$
$$X_5 = [-0.3164 \quad 0.5361]$$

Vector determination using cosine similarity values we rank results in decreasing order. Using

$$Sim(q,d) = \frac{q.d}{|q|\,|d|} \quad . \qquad (6)$$

Hence,

$$Sim(Y, X_1) = 0.516924$$
$$Sim(Y, X_2) = 0.360846$$
$$Sim(Y, X_3) = 0.237192$$
$$Sim(Y, X_4) = 0.929062$$
$$Sim(Y, X_5) = 0.94228$$

From the calculation $X_5 > X_4 > X_1 > X_2 > X_3$ .That means the value may be interpreted as the proportion of variability in Y that is explained by $X_1, X_2, X_3, X_4, X_5$ .It reveals that birth weight is closely related to Haemoglobin absorption. Haemoglobin is one of the vital roles in the birth weight of babies.

A comparison study is done between Singular Value Decomposition and multiple linear regression model to analyze the relationship between birth weight and maternal variable using the linear regression model of the form "Baby weight, Y" is taken as the dependent variable and the other variables are treated as independent variables.

TABLE II : REGRESSION COEFFICIENT

| Predicator | Coefficient | t | VIF |
|---|---|---|---|
| Constant | 3342.8 | 5.68 | |
| $X_1$ | -96.82 | -1.09 | 283.827 |
| $X_2$ | 9.05 | 0.22 | 75.032 |
| $X_3$ | 68.39 | 0.78 | 309.698 |
| $X_4$ | -29.46 | -0.58 | 59.953 |
| $X_5$ | 45.02 | 0.56 | 106.884 |

From the above TABLE II it is observed that the value of $R^2$ is 0.738. It includes the maternal weight, height, age of mothers, blood pressure and Haemoglobin. It seems to 73.8% of the variation in baby weight is explained by the fitted model. The remaining 16.2 of variation can be explained by the factors other than these variables like socioeconomic factors.

Analyzing has been done for each maternal variable with the birth weight as shown below.

TABLE III: $R^2$ VALUES OF ALL INDEPENDENT VARIABLE

| Dependent variable | Independent variable | $R^2$ |
|---|---|---|
| $Y$ | $X_1$ | 7.3 |
| $Y$ | $X_2$ | 1.5 |
| $Y$ | $X_3$ | 2.7 |
| $Y$ | $X_4$ | 3.0 |
| $Y$ | $X_5$ | 8.1 |

From the above Table III shows that the relation between Y and X1 is 7.3%, Y and X2 = 1.5, Y and X3 = 2.7, Y and X4 = 3.0 Y and X5 = 8.1

Hence also hemoglobin concentration is very close to birth weight .Hemoglobin is one of the vital role in the birth weight of babies.

TABLE IV: COMPARISON STUDY

| Variable related to Y | Cosine Similarity | Multiple regression $R^2$ |
|---|---|---|
| $X_1$ | 0.5169 | 7.3 |
| $X_2$ | 0.3608 | 1.5 |
| $X_3$ | 0.2371 | 2.7 |
| $X_4$ | 0.9298 | 3 |
| $X_5$ | 0.94228 | 8.1 |

Above table reveals that birth weight is very close to its haemoglobin concentration

## IV. CONCLUSION

From this study, it is observed that, the birth weight mainly depends on the Haemoglobin concentration in both Sigular Value Decomposition analysis and multiple linear regression analysis. Hence the mother with high Haemoglobin concentration can avoid low birth weight. So the pregnant women should intake additional nutritional food to increase the Haemoglobin concentration and to avoid the health risk problems among the neonates. Some following reasons are the case of lowbirthweight, that is the mother has not obtained the appropriate nutrition, early marriage, late pregnancy at around 35 years, mother below 40 Kilograms, mother has anemia problems, small placenta, chronic placenta insufficiency can also lead to low birth weight , lack of Oxygen also leads low birth weight and due to mothers hypertension and malnutrition. Some action taken before birth to avoid low birth weight, which is regular checkup, appropriate nutrition intake, checks the haemoglobin status, take iron. The case should be identified and corrected in the mother inability to gain weight should be deleted early to avoid problems. The following actions are taken after birth. In such circumstance the child needs extra care to maintain the temperature using like incubator, overhead warmer and kangaroo mother care.

REFERRENCE: